\documentclass[preprint,12pt]{elsarticle}

\usepackage{amssymb}
\usepackage{hyperref}
\usepackage{multirow}

\journal{Journal of Biomedical Signal Processing and Control}

\begin{document}

\begin{frontmatter}



\title{Contrasting Deep Learning Models for Direct Respiratory Insufficiency Detection \textit{Versus}  Blood Oxygen Saturation Estimation}


\author[1]{Marcelo Matheus Gauy\footnote{Corresponding author: marcelo.gauy@usp.br}}
\author[1]{Nat\'alia Hitomi Koza}
\author[1]{Ricardo Mikio Morita}
\author[2]{Gabriel Rocha Stanzione}
\author[3]{Arnaldo C\^andido J\'unior}
\author[4]{Larissa Cristina Berti}
\author[5]{Anna Sara Shafferman Levin}
\author[5]{Ester Cerdeira Sabino}
\author[6]{Flaviane Romani Fernandes Svartman}
\author[1]{Marcelo Finger}

\affiliation[1]{organization={IME - University of São Paulo},
            addressline={Rua do Matão 1010}, 
            city={São Paulo},
            postcode={05508-090}, 
            state={São Paulo},
            country={Brazil}}

\affiliation[2]{organization={Federal Technological University of Parana},
            addressline={Avenida Brasil, 4232}, 
            city={Medianeira},
            postcode={85884-000}, 
            state={Paraná},
            country={Brazil}}

\affiliation[3]{organization={IBILCE - São Paulo State University},
            addressline={Rua Cristóvão Colombo 2265}, 
            city={São José do Rio Preto},
            postcode={15054-000}, 
            state={São Paulo},
            country={Brazil}}

\affiliation[4]{organization={São Paulo State University},
            addressline={Av. Hygino Muzzi Filho, 737}, 
            city={Marília},
            postcode={17525-900}, 
            state={São Paulo},
            country={Brazil}}

\affiliation[5]{organization={FMUSP - University of São Paulo},
            addressline={Av. Dr. Arnaldo, 455}, 
            city={São Paulo},
            postcode={01246-903}, 
            state={São Paulo},
            country={Brazil}}

\affiliation[6]{organization={FFLCH - University of São Paulo},
            addressline={Rua do Lago, 717}, 
            city={São Paulo},
            postcode={05508-080}, 
            state={São Paulo},
            country={Brazil}}

\begin{abstract}

We contrast high effectiveness of state of the art deep learning architectures designed for general audio classification tasks, refined for respiratory insufficiency (RI) detection and blood oxygen saturation (SpO2) estimation and classification through automated audio analysis. 
Recently, multiple deep learning architectures have been proposed to detect RI in COVID patients through audio analysis, achieving accuracy above $95\%$ and F1-score above $0.93$.
RI is a condition associated with low SpO2 levels, commonly defined as the threshold SpO2 $<92\%$. While SpO2 serves as a crucial determinant of RI, a medical doctor's diagnosis typically relies on multiple factors. These include respiratory frequency, heart rate, SpO2 levels, among others. 
Here we study pretrained audio neural networks (CNN6, CNN10 and CNN14) and the Masked Autoencoder (Audio-MAE) for RI detection, where these models achieve near perfect accuracy, surpassing previous results. Yet, for the regression task of estimating SpO2 levels, the models achieve root mean square error values exceeding the accepted clinical range of $3.5\%$ for finger oximeters. Additionally, Pearson correlation coefficients fail to surpass $0.3$. As deep learning models perform better in classification than regression, we transform SpO2-regression into a SpO2-threshold binary classification problem, with a threshold of $92\%$. However, this task still yields an F1-score below $0.65$. Thus, audio analysis offers valuable insights into a patient's RI status, but does not provide accurate information about actual SpO2 levels, indicating a separation of domains in which voice and speech biomarkers may and may not be useful in medical diagnostics under current technologies.

\end{abstract}

\begin{graphicalabstract}
\includegraphics[width=1\textwidth]{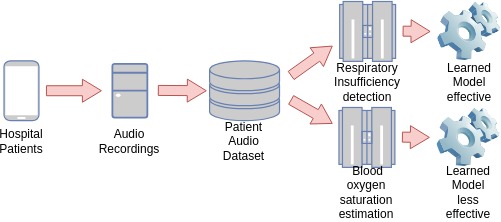}
\end{graphicalabstract}

\begin{highlights}
\item We improve literature results on audio-based respiratory insufficiency detection.
\item Blood oxygen saturation estimation from audio is hard under current technologies.
\item We analyze features making the former effective but not the latter.
\end{highlights}


\end{frontmatter}



\section{Introduction}

Respiratory insufficiency (RI) is a condition commonly associated with low blood oxygen saturation levels (SpO2), which can be broadly defined as the impairment of respiratory gas exchange between the ambient air and circulating blood~\citep{irwin2004cardiopulmonary}. Interest in its automated detection has intensified during the pandemic, as it is a COVID-19 associated condition that leads to hospitalization~\citep{tobin2020covid}.  However, for the purposes of this study, RI is defined as blood oxygen saturation level below a certain threshold, most usually  SpO2 $<92\%$~\citep{spira2021}. 

It's important to acknowledge that there is no universally definitive gold standard for diagnosing respiratory insufficiency. Medical practitioners typically integrate various factors into their diagnostic process, including respiratory frequency, heart rate, and SpO2 levels, among others. Consequently, in clinical practice, scenarios arise where patients may exhibit SpO2 levels potentially above the threshold, yet other clinical indicators point towards a positive diagnosis of RI. Conversely, instances occur where SpO2 levels fall below the threshold, but other clinical factors do not support a diagnosis of RI. An extreme example of the latter could be measuring the SpO2 of a swimmer immediately following an extensive swimming exercise.

Recent works analyze audios of COVID-19 induced RI patients~\citep{spira2021, 219270} as well as general RI~\citep{aime2023gauy} and determined that deep learning architectures --- such as Convolutional Neural Networks, or CNNs, and Transformers based architectures --- are very effective at distinguishing audio recordings of RI patients from non-RI control voices. The most effective architectures achieve accuracy between $95\%$ and $97.5\%$ on either COVID-19 induced RI dataset or general RI dataset~\citep{aime2023gauy}. Factors believed to be important in the decision process of those models are altered speech pause distribution in patients~\citep{fernandessvartman22_speechprosody} as well as signal energy levels throughout speech~\citep{spirainterpretability2022} and altered $F_0$ related parameters~\citep{berti2023fundamental}. These analyses have been carried out as part of the SPIRA project~\citep{SpiraAccoustic2021, SPIRA-PMLD2022}, which aims to provide cheap and mobile artificial intelligence (AI) tools for the triage of RI patients via audio analysis.

In this paper, we study current state of the art (SOTA) audio classification models and their proficiency for detecting RI. The models studied are the Pretrained Audio Neural Networks (PANNs~\cite{kong2020panns}) in addition to a novel unsupervised pretrained model called Masked Autoencoder (Audio-MAE~\cite{huang2022masked}). These models are pretrained on a $5000$ hours dataset of Youtube videos called AudioSet~\cite{gemmeke2017audio}, and are known to be extremely effective at a large array of audio classification tasks~\cite{kong2020panns, huang2022masked}. Moreover, while RI detection has been extensively studied, no previous works have analyzed whether it is possible to directly estimate SpO2 (a key defining characteristic of RI) from voice  and speech\footnote{We distinguish between the analysis of \textit{voice}, namely acoustic properties of human utterances, from \textit{speech}, which involves natural language emissions.}. Thus, we investigate whether the previously mentioned SOTA audio classification models can be used for estimating patient SpO2 solely from voice and speech audios. The architectures considered are extremely effective at RI detection, and surpass previous works, with the best model for RI detection being the Audio-MAE, achieving near perfect accuracy ($99.9\%$). However, despite the significant improvement in RI detection, as well as their widespread effectiveness at multiple audio classification tasks, the SOTA models studied are not capable of attaining good performance when estimating SpO2. The $4$ networks are used in a regression task of estimating patient SpO2 from a patient's voice and speech audios. Moreover, as a regression task may be challenging for the networks due to the relatively small size of the patient dataset, we also use them in a quantized classification task performed over classes of low SpO2 and high SpO2 for a given threshold, e.g. $92\%$.

For the regression task, no model achieves Pearson correlation between the oximeter SpO2 values and the predicted SpO2 values above $0.3$, with the mean absolute error being around $4.4\%$ for the best model~\footnote{It is known that oximeter error, depending on the instrument, varies between $1\%$ and $2\%$.}.
For the classification task, all models display F1-score below $0.65$. Note that this is significantly lower than the result obtained for RI detection, as the F1-score would also be above $0.99$.

Thus, this paper presents the following contributions:
\begin{itemize}
\item Audio-MAE and the PANNs improve the RI detection accuracy to near perfection, significantly surpassing previous models.
\item Despite that, none of the studied models is capable of estimating SpO2 accurately through voice and speech. This showcases the potential limits of using both as biomarkers.
\item We present a list of reasons why models might be able to detect RI but struggle with SpO2.
\end{itemize}




\section{Related Work}

This work is developed as part of the SPIRA project~\citep{SpiraAccoustic2021, SPIRA-PMLD2022}. Previous works within that context have built effective models for the detection via audio analysis (voice and speech) of both COVID-19 induced RI~\citep{spira2021, 219270, gauy2023acoustic}, and general RI~\citep{aime2023gauy}, where the detection is aimed at Brazilian Portuguese speech. Variants of some of those models were also proposed for the detection of COVID-19 through voice~\citep{casanova21_interspeech, spirainterpretability2022}. In addition to the development of deep learning models, the SPIRA project also aims at studying the acoustic properties of RI compromised speech, that is, it aims at investigating the characteristics present in voice and speech which support those models to make their classification, thus aiming to generate explanations for its decisions. It has been found that RI patients have altered speech pause distribution~\citep{fernandessvartman22_speechprosody}, different $F_0$ related parameters~\citep{berti2023fundamental} as well as distinct signal energy throughout speech~\citep{spirainterpretability2022}. Thus, previous works from the SPIRA project focused on analyzing RI and its effects on voice and speech directly, either through building deep learning models or through determining explainable features. No previous analysis from the project focused on analyzing SpO2 levels directly and estimating those from voice and speech.

Outside the SPIRA project, we have not found works studying the relations between RI and voice or speech. However, there are initiatives which try to detect COVID-19 from voice or cough~\citep{pinkas2020sars, laguarta2020covid, despotovic2021detection, watase2023severity}, as well as recent works studying COVID-19 disease progression via longitudinal cough, breathing and voice data~\citep{dang2022exploring}.  We must also mention that~\cite{kaufman2023acoustic} have recently proposed employing acoustic analysis for the detection of type-2 diabetes, but further studies seem  necessary to validate this approach.

While no previous works studying the problem of determining SpO2 by means of voice and speech analysis have been found, it is common to use machine learning techniques to convert the raw PPG signal from an oximeter into the SpO2 values~\citep{shuzan2023machine, venkat2019machine, priem2020clinical}. This process is reported to have a mean absolute error below $1\%$ and an above $95\%$ accuracy for a $\pm 2\%$ error band for SpO2 values in the range $81$ to $100$.

\section{Methods}

\subsection{SPIRA Dataset}
\label{sec:dataset}

Here, we use the original dataset from~\citep{spira2021, 219270} for RI detection and an extended version of the corresponding patient dataset for SpO2 estimation. The original dataset consisted of $566$ patient audios (and about $6000$ control audios). These audios mainly consisted in patient's uttering a predetermined sentence which they would read from a prompt on a cellphone \footnote{A sentence was designed by linguists to contain simple but large words, as well as no obvious speech pauses during its utterance, leading to: ``O amor ao pr\'oximo ajuda a enfrentar o coronav\'irus com a for\c{c}a que a gente precisa''/\textit{``Love of neighbor helps in strengthening the fight against Coronavirus.''}}. These patient voices were collected in COVID-19 wards in Brazil during the peak of the COVID-19 pandemic. The control audios were collected from voluntary donors with an app over the internet. Collection was absolutely anonymous, so no one knows who the patients and controls were, and no ethnographic information is available. More than 600 patient audios were collected, but a number of distortions, most commonly the whispering of collectors being heard in the audios, led to several audios being discarded, leading to a $566$ item dataset~\citep{spira2021}. 

In this work, we use the exact same dataset as~\cite{spira2021} with the same division in training, validation and test sets proposed there. This is to allow comparisons between our models and previous studies. For SpO2 estimation, we use the patient audios (as they are the ones with SpO2 available) and, in this case, do not need to contrast them with the control audios (collected in another environment). We can use the full original patient dataset for SpO2 estimation. In total, we have $566$ patient audios with their respective SpO2 measured with an oximeter at the time of voice recording. Alternatively, we could have used the more general RI dataset provided by~\citep{aime2023gauy}, which includes SpO2 data for both patients and controls. However, that dataset contains less than $200$ files and is too small to use on its own in a regression task. In addition, it may not be a good idea to combine both datasets as the voice features from COVID-19 RI and general RI are quite different~\citep{aime2023gauy}. As such, we found it better to use the COVID-19 induced RI audios we had available. As it might still be possible to identify patient's by their voice, and as healthcare datasets are generally confidential, we prefer to not publish the entirety of our patient dataset. Note that the dataset from~\citep{spira2021} is found on~\href{https://github.com/SPIRA-COVID19/SPIRA-ACL2021/tree/master}{Github}.

\begin{figure}[tb]
\centering
\includegraphics[width=0.7\textwidth]{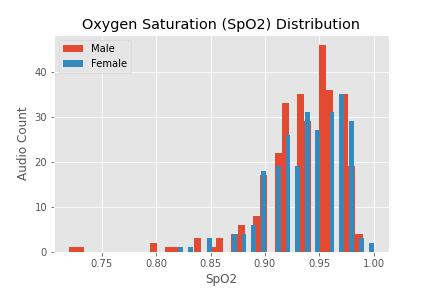}
\caption{SpO2 distribution. Men's SpO2 mean  is $93.4$. For women it is $94.0$.} \label{figure:oxygen_saturation_distribution}
\end{figure}

In the full $566$ audios dataset, we have $307$ men, of which $103$ have SpO2 below or equal to $92\%$, and $259$ women, of which $82$ have SpO2 below or equal to $92\%$. We perform a random split of the dataset in training, validation and test sets in the following way: first, among the $103$ men and $82$ women with SpO2 below or equal to $92\%$, randomly select $8$ men and $8$ women for the validation set and $16$ men and $16$ women for the test set, the remaining files with SpO2 below or equal to $92\%$ go to the training set; second, among the $204$ men and $177$ women with SpO2 above $92\%$, randomly select $16$ men and $16$ women for the validation set and $32$ men and $32$ women for the test set, with the remaining files with SpO2 above $92\%$ going to the training set. This is to ensure that our validation and test sets are balanced by sex and have similar proportions of low and high SpO2 present between the three dataset parts.

\subsection{RI detection task} 

As mentioned previously, we use the exact same training, validation and test sets division as the original work~\cite{spira2021}. In the original dataset, patients who suffered from RI were collected in 
COVID-19 hospital wards and controls were collected via a web app. The different collection environments require that the audios are properly preprocessed so as to avoid overfitting to the different noise sources~\citep{spira2021}. We perform the same preprocessing steps for RI detection as suggested by previous works~\citep{spira2021, 219270, aime2023gauy}. Our models receive spectrograms as opposed to MFCC-grams as was common in past studies. This is due to their effective pretraining on AudioSet, which, through transfer learning, allows the models to perform extremely well despite using spectrograms (which were found to be typically less effective than MFCC-grams in previous works).

\subsection{SpO2 Estimation tasks}

We perform two types of SpO2 estimation tasks. The first is a regression task where the models predict a value between $0$ and $100$ that should match as closely as possible the SpO2 level of the corresponding patient. We use the MSE (mean squared error) loss for gradient computation, though we also made preliminary experiments with the MAE (mean absolute error) loss obtaining similar results for the models we tested. The second task is a classification one where the models are asked to classify whether the audios come from a patient with SpO2 above $92\%$ or not. This is a binary classification task that is likely to be much easier for the models than performing a complete SpO2 regression and also likely to be much less data intensive, which is important as healthcare datasets are usually small in size. This allows us to check whether the apparent difficulty from the regression task did not come solely from the size of the dataset. The loss used is a binary cross entropy loss, as is standard for binary classification.

Lastly, we have observed, especially for regression, that there is some reasonably large performance difference (measured in terms of the Pearson coefficient for regression) among different training, validation and test splits. This did not seem significant in our preliminary experiments for the classification task but did seem significant for regression. To more accurately represent the performance of the models, we repeat each experiment on each model $10$ times with the added caveat that each experiment gets a different (random) training, validation, test set split (as proposed in Section~\ref{sec:dataset}). Thus, our average performance (say, Pearson coefficient for regression or F1-score for classification) on the test set more closely resembles the average one from a random test set and not the particular test set for any of the experiments. Alternatively, we could have used cross-validation to avoid the problem of selecting the test set, but that would often require changing the internal logic of the data feeding for each model as the windowing augmentation technique we use needs to be performed after the fold division (see Section~\ref{section:preprocessing} for more details). As a result, we chose to just resample the test set with each experiment and get an average result.

\subsection{Model Architectures}
\label{section:model_architectures}

\begin{figure}[tb]
\centering
\includegraphics[width=0.9\textwidth]{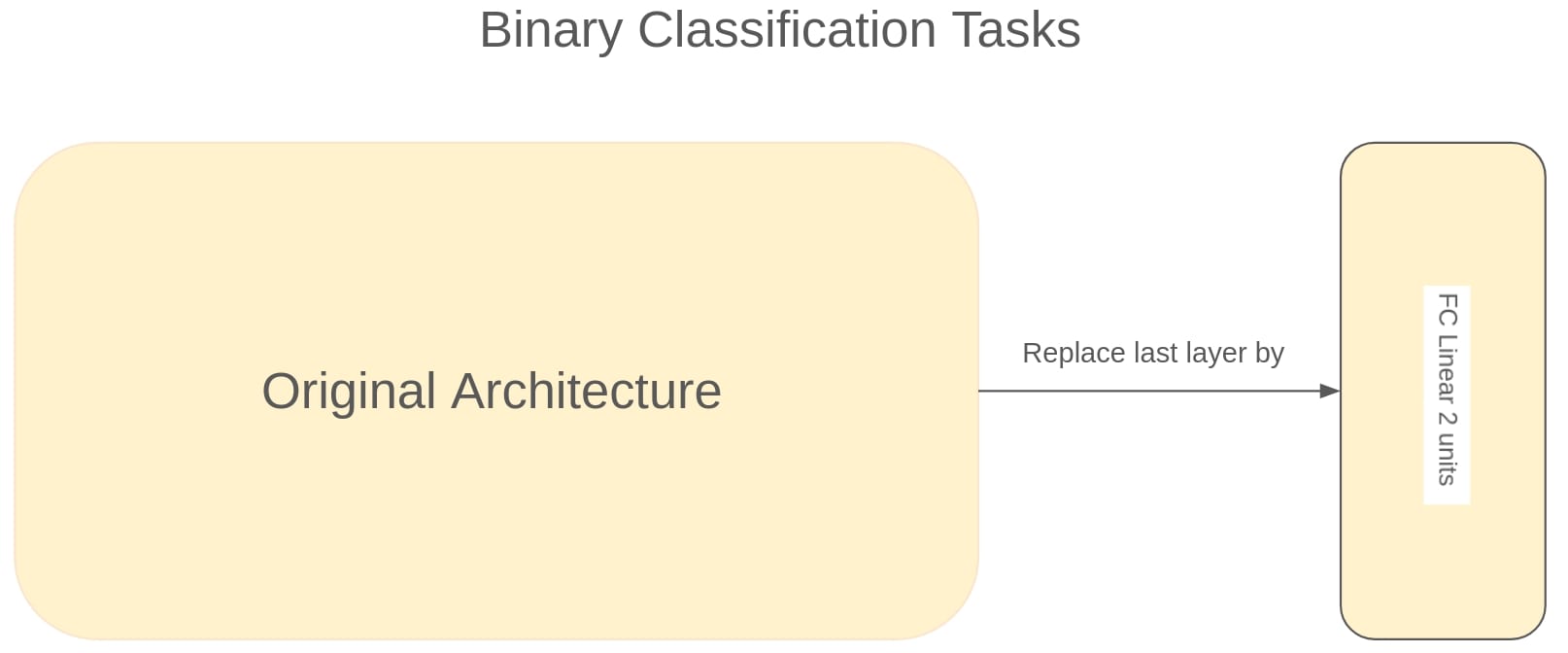}
\caption{Binary classification task architecture structure. Original Architecture refers to either Audio-MAE or the PANNs (CNN6, CNN10, CNN14). FC Linear 2 units is a fully connected (FC) linear layer with $2$ units to which we use softmax as part of the BCEwithLogits loss. This architecture is used for SpO2 classification and RI detection tasks.} \label{figure:classification_architecture}
\end{figure}

\begin{figure}[tb]
\centering
\includegraphics[width=0.9\textwidth]{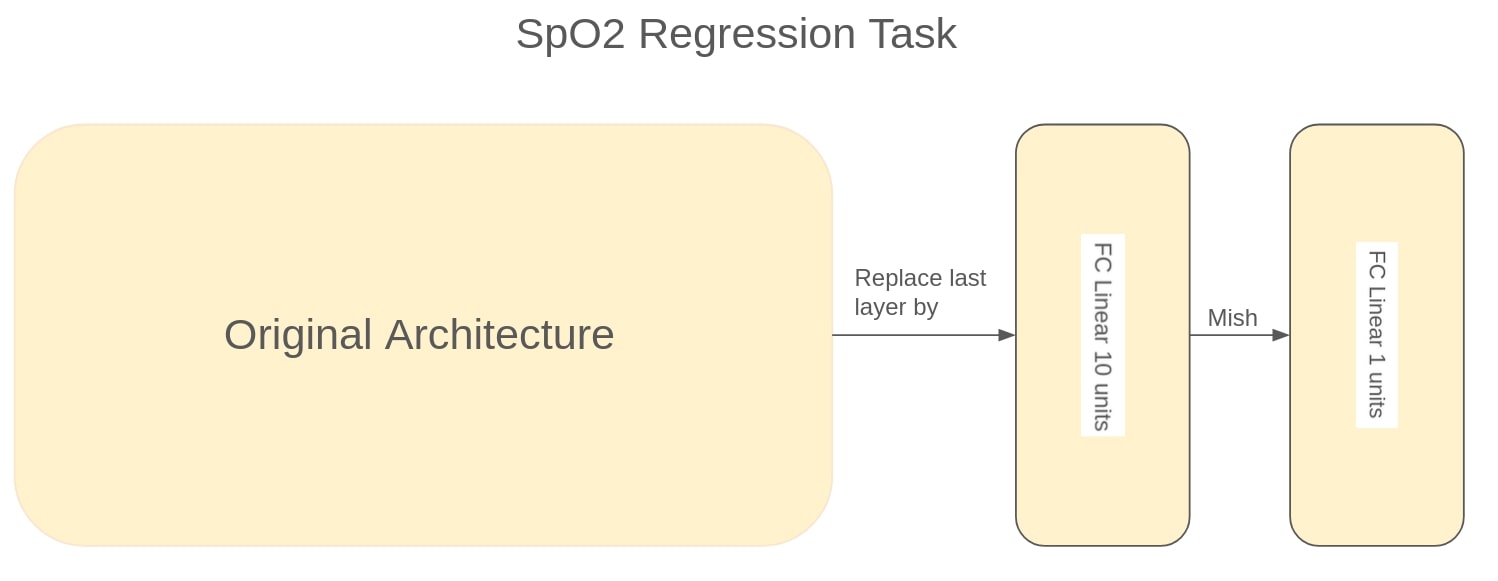}
\caption{SpO2 regression task architecture structure. Original Architecture refers to either Audio-MAE or the PANNs (CNN6, CNN10, CNN14). FC Linear $x$ units is a fully connected (FC) linear layer with $x$ units. We apply the Mish activation function to the intermediary layer. Observe that we have attempted varying the number of units in the intermediary layer between $10, 25, 50, 100$ as well as including dropout between the intermediary layer and the last layer. We also ran experiments using Gelu in place of Mish.} \label{figure:regression_architecture}
\end{figure}

In general, the models described here were used exactly as originally proposed, with the exception of the last layer, which consists of $1$ unit for regression and $2$ units for classification (see Figure~\ref{figure:classification_architecture} for the general classification architecture). In the case of regression, we add an extra $10$ units fully-connected feed-forward (FC) layer before the final layer to facilitate the information extraction. See Figure~\ref{figure:regression_architecture} for the general SpO2 regression architecture. We find that applying dropout at the intermediary FC layer is not helpful (so increasing its size and regularizing with dropout does not lead to improved results). The codes used are the same as the original models with the few changes mentioned in Figures~\ref{figure:classification_architecture} and~\ref{figure:regression_architecture}.

\textbf{Masked Autoencoder (Audio MAE)}
This model was proposed in~\citep{huang2022masked}. Similar variants of this approach can be found in~\citep{gong2022ssast, baade2022mae, niizumi2022masked, chong2023masked}. Masked Autoencoders were originally proposed for image processing~\citep{he2022masked} and later found to also perform extremely well in audio classification tasks~\citep{huang2022masked}. The input to the Masked Autoencoder is typically the spectrogram of the audio. The key idea here is to do masked reconstruction of the frames, an idea inspired by the pretraining technique of~\citep{devlin2018bert} for NLP. The authors propose to erase a large proportion of the audios ($80\%$ in this case). Additionally, the authors do not erase frames per se but typically $16\times 16$ blocks of frames and channels, borrowing from Vision Transformers~\citep{dosovitskiy2021an} method of feeding the spectrogram in blocks as tokens to the Transformers. The models are (unsupervisedly) pretrained to reconstruct this large share of masked blocks back into the original spectrogram. In order to deal with a large masking proportion, the authors make use of a standard Transformer both for encoding and decoding. The encoder is a vanilla $12$-layer Vision Transformers~\citep{dosovitskiy2021an}. The decoder is a $16$-layer Transformers with shifted local attention~\footnote{This is a variant of local attention which shifts the attention windows from one layer to the next. The reason for doing this instead of standard global attention, as is typically used for images, is that the relevant information in audio spectrograms is predominantly local.}.

We use the pretrained Audio-MAE model on AudioSet from~\citep{huang2022masked}. AudioSet is a $5000$ hour dataset of Youtube audios distributed in $527$ classes. In principle, one could consider the version of this model also finetuned (that is, trained with supervision) on the AudioSet. However, as we deal with a regression task, we did not expect there to be a significant advantage from doing this type of finetuning on top of the pretraining~\footnote{Moreover, the Pretrained Audio Neural Networks described below were trained on AudioSet with supervision, so we already explore whether that has some inherent advantage.}. As such, we have settled for the model that received unsupervised pretraining on AudioSet.
Lastly, in our experiments with the COVID-19 induced RI detection task (dataset from~\citep{spira2021}), the accuracy of the model is above $99.9\%$ and even in the harder general RI detection task (dataset from~\citep{aime2023gauy}) the accuracy is above $98\%$, without employing any changes to Audio-MAE. This showcases the effectiveness of the pretraining technique used as previous models had lower accuracy and typically only achieved it with MFCC-gram as the input.

\textbf{Pretrained Audio Neural Networks - PANNs}
We make use of $3$ CNN-based PANNs from~\citep{kong2020panns}, namely CNN6, CNN10 and CNN14. These are pretrained (with supervision) on the AudioSet dataset to classify the audios among the $527$ classes. They have been effectively used for multiple audio based tasks, such as, audio set tagging~\citep{kong2020panns}, speech emotion recognition~\citep{gauy2022pretrained}, automated audio captioning~\citep{xu2021investigating} 
and COVID-19 detection~\citep{spirainterpretability2022}. These $3$ PANNs are similar in architecture but have varying levels of complexity. We find this useful as it gives us an idea on how hindered our models are by the size of our dataset.

CNN6 is a $6$ layer CNN; CNN10 has $10$ layers; lastly CNN14 has $14$ layers. They have convolutional layers, with kernel $5\times 5$ for CNN6, and kernel $3\times 3$ for CNN10 and CNN14. Each such layer is followed by batch normalization~\citep{ioffe2015batch} and ReLu nonlinearity~\citep{nair2010rectified}, for faster and more stable training convergence~\citep{kong2020panns}. These layers are present $4$ times in CNN6 and, in between two layers, an average pooling $2\times 2$ (AVG) is used, which works better than max pooling~\citep{kong2019cross}. For CNN10 and CNN14, the convolutional layers appear in pairs before AVG is applied. CNN10 contains $4$ pairs of such blocks, for a total of $8$ layers, and CNN14 contains $6$ pairs of such blocks, for a total of $12$ layers. As for the number of kernels in each block pair, they start at $64$ for the first pair and double for each subsequent pair. In the last block pair, instead of AVG, we apply Global pooling (sum of average and max pooling). All networks have a penultimate FC layer to increase representation ability, containing $512$ units on CNN6 and CNN10 and $2048$ units on CNN14; and a final $527$ unit FC sigmoid layer, to obtain the class probabilities. To prevent overfitting, dropout is applied after every downsampling operation. All networks take as input a sequence of frames of log mel Spectrogram with $64$ mel bins each. 

We use the pretrained models from~\citep{kong2020panns}. These were pretrained to classify the AudioSet audios in $527$ classes. 
Finally, the $3$ pretrained PANNs are also superior to the MFCC Transformers on both standard RI detection tasks (COVID-19 induced RI and general RI), so they are known to be effective for RI detection~\citep{aime2023gauy}.

\subsection{Preprocessing}
\label{section:preprocessing}

We perform the same windowing data augmentation technique used in~\citep{spira2021}. Namely, our audios are split in $4$ second windows with $1$ second hop. That is, a $5$ second audio is split in two, one from seconds $0$ to $4$ and the other from seconds $1$ to $5$. This is a simple data augmentation that increases the amount of audios we have available. Moreover, as Audio-MAE uses all-to-all attention, decreasing the length of the audios leads it to be less computationally and memory intensive. Furthermore, there may be some correlation between lower SpO2 levels and audio length, as patients with lower SpO2 may have more difficulty breathing (as was observed in previous works when contrasting healthy controls and RI patients~\citep{fernandessvartman22_speechprosody}), and we want to prevent the model from excessively concentrating on the audio lengths. Lastly, observe that windowing needs to be performed after the training, validation and test set split as otherwise we would risk biasing our results as the validation and test sets would potentially share a lot of audio parts with the training set.

After the windowing step is performed, we resample the audio waveforms with a $16kHz$ sample rate~\footnote{Performance difference by resampling the audios is minimal.} and convert them into either a spectrogram, via a Fast Fourier transform (FFT)~\citep{brigham1967fast}, feeding them to the models as an image. The parameters for conversion (such as, hop length, window length, number of mel coefficients and size of FFT) vary with the model as we are typically using pretrained versions of models, which were pretrained with a particular set of parameters. Apart from being required to use the same parameters when extracting the spectrogram as the original pretrained models, we do not expect there to be major differences in performance due to the particular choice of those parameters as long as they are within reason.

Specifically for the COVID-19 RI detection task, one needs to add hospital ward noises to both patient and control audios. This is the same process as in previous works~\citep{spira2021, 219270, gauy2023acoustic} and is a necessary step to avoid overfitting to such noises. This step is after windowing, but naturally before spectrogram transformation.

\section{Results}
\label{section:results}

\subsection{COVID-19 RI detection task}

We performed an experiment to evaluate the accuracy of the $4$ SOTA models studied here in the COVID-19 RI detection task. We also briefly report the same results for the general RI detection task using the dataset from~\citep{aime2023gauy}. We use the binary cross entropy loss to train the $4$ models described in Section~\ref{section:model_architectures} for COVID-19 RI detection on the training set from~\cite{spira2021}, performing early stopping according to the best accuracy measured on the validation set at the end of each epoch. All models were run with a batch size of $16$, with Adam optimizer~\citep{kingma2014adam} with the same parameters as the original works for the SOTA models. The learning rate for the PANNs was $0.0001$ without any scheduler and was trained, with early stopping, for $100$ epochs. Audio-MAE used the original learning rate with the original Noam scheduler and was trained, with early stopping, for $40$ epochs. Experiments with PANNs were repeated $5$ times, while the Audio-MAE experiment was repeated $10$ times to obtain an average behavior. We only report accuracy as the dataset is balanced between patients and controls as well as sex. We also report the accuracy on the dataset from~\cite{aime2023gauy}. Since that is a smaller dataset, the Audio-MAE performance has larger variance, which is probably a consequence of some overfitting in specific runs. Perhaps this could be avoided with hyperparameter optimization but it is nevertheless a complex model on a tiny dataset. The results can be found on Table~\ref{table:ri_detection}.

\begin{table}
\centering
\caption{RI detection task. Note that the previous best model on the dataset~\cite{spira2021} had accuracies below $97.5\%$~\citep{gauy2023acoustic} and around $95\%$ for the general RI dataset from~\cite{aime2023gauy}. We report sample mean and sample standard deviation across $5$ experiments for CNN6, CNN10 and CNN14 and $10$ experiments for Audio-MAE.}\label{table:ri_detection}
\begin{tabular}{|c | c| c|}
\hline
Model  &  Accuracy on dataset~\cite{spira2021} &  Accuracy on dataset~\cite{aime2023gauy} \\ \hline
Audio-MAE & $\textbf{99.98}\%\pm 0.01$ & $98.30\%\pm 2.04$   \\ \hline 
CNN6  & $97.84\%\pm 1.05$ & $\textbf{98.51}\%\pm 0.62$\\ \hline
CNN10  & $98.29\%\pm 1.20$ & $97.66\%\pm 0.42$      \\ \hline
CNN14  & $97.93\%\pm 1.14$ & $97.66\%\pm 0.64$      \\ \hline

\end{tabular}
\end{table}

\subsection{SpO2 estimation analysis}

We perform two experiments to showcase that voice and speech features do not contain significant information regarding a patient's SpO2 level. The first experiment is a direct SpO2 regression task, where models are tasked with predicting the SpO2 level of a patient from audio samples. We use the MSE loss to train the $4$ models described in Section~\ref{section:model_architectures} on the training set, performing early stopping according to the best MSE loss measured on the validation set at the end of each epoch. We evaluate the models according to $4$ metrics: root mean squared error (RMSE), MAE, $R^2$ and Pearson correlation. As we observe large variance in performance depending on the run and dataset split, we perform $10$ random dataset splits into training, validation and test sets and perform train-test cycles $10$ times. All models were run with a batch size of $16$. MOdels were trained with AdamW optimizer~\citep{loshchilov2017decoupled}, under default weight decay of $0.01$, except for the Audio-MAE, which used the original value of $0.0005$.
PANNs used a learning rate of $0.0001$ without any scheduler. They were trained for $10$ epochs. The Audio-MAE used the original initial learning rate with the original Noam scheduler parameters and was trained for $20$ epochs.

The $4$ models used the originally proposed hyperparameter sets and we then performed a search around those, going one by one and searching for the best value while fixing all others. No significant advantage was observed by varying parameters.

Table~\ref{table:regression} shows the results of each model according to the $4$ metrics. The average across $10$ experiments with the sample standard deviation is depicted. No model achieved Pearson above $0.3$ and $R^2$ above $0.1$ on average, showcasing the difficulty of the task and the seemingly little information on the SpO2 level available on a patient's voice. Models typically stayed in the range of $0.22$ and $0.26$ for the average Pearson and, considering the reported sample standard deviation, they can all be considered almost equivalent to one another, except for CNN14.

\begin{table}
\centering
\caption{SpO2 Regression Task on Patient Dataset. We report average RMSE, MAE, $R^2$ and Pearson across $10$ experiments along with the sample standard deviation.}\label{table:regression}
\begin{tabular}{|c | c| c| c| c|}
\hline
Model & RMSE &  MAE & $R^2$ & Pearson \\ \hline

Audio-MAE & $\textbf{4.4}\pm 0.8$  & $\textbf{3.0}\pm 0.4$          & $-0.135\pm 0.06$ & $0.233\pm 0.083$  \\ \hline
CNN6  & $4.8\pm 1.0$           & $3.9\pm 0.8$ & $0.069\pm 0.05$ & $0.227\pm 0.130$ \\ \hline
CNN10 & $4.5\pm 0.7$  & $3.4\pm 0.6$          & $\textbf{0.074}\pm 0.05$ & $\textbf{0.251}\pm 0.108$ \\ \hline
CNN14 &  $4.7\pm 0.8$          & $3.7\pm 0.6$          & $0.004\pm 0.01$ & $0.034\pm 0.055$ \\ \hline

\end{tabular}
\end{table}

As the previous regression task may have been too hard for the models given the amount of data we had available for training, we proposed a second experiment to support our result that the voice of patient's does not contain much information on their SpO2 level. For the second experiment we reduce the task of estimating SpO2 levels to a simple binary classification task: whether the SpO2 of a patient is above a given threshold or not. We have chosen the threshold at $92\%$ as that is commonly the defining SpO2 level for an RI patient. With this threshold, we can see the task as distinguishing between more critical RI patients (those still following the textbook definition for RI based on SpO2) and less critical ones, where criticality depends solely on the SpO2 level. Note that considering only SpO2 is likely ill-advised when defining criticality as the data comes from COVID-19 wards. However, this task does allow us to measure whether voice features contain information on the SpO2 of a patient, or whether the information they contain refers more generally to a patient's condition and has little influence from the SpO2.

We use a binary cross-entropy loss to train the $4$ models described in Section~\ref{section:model_architectures} on the training set and perform early stopping according to the best loss measured on the validation set at the end of each epoch.
We evaluate the models according to their classification accuracy and more importantly their F1-score. The F1-score is effectively the correct measure to use as the number of patients with SpO2 above $92$ is much larger than the number of patients with SpO2 below or equal to $92$.

As in the regression task, the variance is quite large among experiments. Though the influence of the dataset split is smaller, for symmetry with the regression task, we consider the same $10$ dataset splits into training, validation and test sets and perform $10$ experiments on each model, one on each dataset split. This gives us an average performance of the models for a random test set. All models were run with a batch size of $16$. All models used Adam as the optimizer with the default weight decay of $0.01$, with the exception of the Audio-MAE which used AdamW with a weight decay of $0.0005$ as was the case in the original paper.
The PANNs used a fixed learning rate of $0.0001$ as was the case for regression and were trained for $10$ epochs. The Audio-MAE used the original learning rate and original Noam scheduler parameter and was trained for $40$ epochs. For the second task it was not helpful to have an intermediary layer in the models. Moreover, changing dropout in the inner layers of the pretrained models is likely ill-advised as it would require pretraining to be done again. We have attempted, without success, to vary the weight decay parameter to increase $L_2$ regularization. It does not seem that one can improve the results substantially by simply exploring the parameter space and to get better results, one would likely need substantially more data. Even then, it seems unlikely that this task can ever reach as high accuracy and F1-score as the RI detection task.

Table~\ref{table:classification} shows the results of each model according to accuracy and F1-score. The average across $10$ experiments is shown with the sample standard deviation. No model achieved an F1-score above $0.65$ on average, which is considerably lower than the above-mentioned more than $0.97$ F1-score achieved by these models on the RI detection task. Due to the overlapping intervals given by mean standard deviation, models based on Masked Autoencoders and the $3$ PANNs can be considered almost equivalent in performance.

\begin{table}
\centering
\caption{SpO2 Classification Task on SPIRA Patient Dataset. We report F1-score and accuracy sample mean and standard deviation across $10$ experiments on each model.}\label{table:classification}
\begin{tabular}{|c | c| c|}
\hline
Model & F1-score  & Accuracy \\ \hline
Audio-MAE & $0.621\pm 0.052$          & $65.9\%\pm 4.72$     \\ \hline
CNN6  & $0.641\pm 0.041$           & $66.34\%\pm 3.86$ \\ \hline
CNN10 & $0.615\pm 0.067$  & $66.30\%\pm 5.60$      \\ \hline
CNN14 &  $\textbf{0.643} \pm 0.045$          & $\textbf{66.66}\%\pm 4.95$      \\ \hline

\end{tabular}
\end{table}

\section{Discussion}
\label{section:discussion}

With the three experiments, we show that it is harder to estimate SpO2 levels based solely on a patient's voice, in comparison to RI detection. The chosen models are close to state of the art for a large array of audio classification tasks (see~\citep{huang2022masked, kong2020panns}). Moreover, all $4$ models are known to be very effective at the related RI detection task. We observe that we have attempted, without success, to also use the original models for RI detection from~\cite{spira2021, gauy2023acoustic} for SpO2 estimation. This suggests that a patient's voice and speech has plenty of features pertaining to whether one suffers from RI, even though it has little information on the patient's SpO2 at the time of collection; so important information is lost in the SpO2 estimation process which is nonetheless used in the classification process. We also hypothesize that the patient's treatment in the hospitals leads to an improvement of SpO2 levels (which seemingly pushes them out of the RI range), but their voice and speech retains multiple traces found in RI patients, so that the models cannot distinguish lower SpO2 from higher SpO2 in the investigated models.

Note that for the regression task, the obtained levels of RMSE for all the models were above the clinical level of $3.5$~\citep{priem2020clinical}. The Pearson correlation coefficient is below $0.3$, as well as the $R^2$ being below $0.1$ imply that there is little correlation between the predicted SpO2 and the oximeter SpO2, making the models very poor SpO2 predictors. For the classification task, the F1-score being below $0.65$ is a strong indication of the difficulty of the task and specially when contrasted with the high (larger than $0.98$) F1-score attained in the RI detection task. Thus, there are multiple features present in voice and speech which enable RI detection but cannot be used for estimating SpO2, even if the task is as simple as identifying SpO2 above or below $92\%$.
Such features are, potentially, the patient's more numerous pauses at unexpected locations hypothesized in~\citep{fernandessvartman22_speechprosody}. Another potential feature would be different $F_0$ related parameters~\citep{berti2023fundamental} as well as different energy levels throughout speech~\citep{spirainterpretability2022}. Note that none of these features observed to be relevant for RI detection can correlate strongly to SpO2 levels (as the networks would learn such simple features) even if they correlate with RI. Much like medical doctors are aware of the fact that SpO2 cannot be used as the sole determining factor of RI, the same holds for deep learning models.

In terms of criticism, our work could be contested on the fact that we did not use Wav2Vec~\citep{baevski2020wav2vec} type networks. However, Wav2Vec main advantages lie on transcription tasks, and Masked Autoencoders typically outperform them for standard classification tasks~\citep{huang2022masked}. Another important criticism is that our data consists solely of hospitalized COVID-19 patients. While it is possible that the results would change for other respiratory conditions, we currently possess no means of verifying that hypothesis as the only dataset available with SpO2 values for multiple types of diseases is too small in size~\citep{aime2023gauy} (the total number of cases with SpO2 below $92$ is less than $10$).
Lastly, it could be said that our dataset is just too small for the networks to learn to identify SpO2. While the dataset leans on the small size for complex models, even the simpler models (e.g. CNN6 and CNN10) failed to extract much meaningful information, while all these models could identify RI with a similar dataset.

\section{Conclusions}

We analyzed the performance of $4$ models, which we show to be superior to previous state of the art RI detection through voice. The best model, Audio-MAE, achieves near perfect accuracy on COVID-19 RI detection ($99.9\%$). In terms of general respiratory insufficiency detection (dataset from~\cite{aime2023gauy}), all $4$ models achieve accuracy above $97\%$. In addition, as these models are known to perform extremely well for a large array of audio classification tasks, we have decided to analyze their performance on two SpO2 estimation tasks from voice.

The first task is SpO2 regression from voice, where the best RMSE is above $3.5$ which is the standard for clinical use of SpO2. Even more crucially, Pearson correlation between oximeter SpO2 values and even the best model SpO2 estimate does not reach $0.3$. Then we considered a binary classification task for predicting from voice whether the SpO2 is above $92\%$ or not. None of the considered models achieve an F1-score above $0.65$. This is in sharp contrast with RI detection, be it COVID-19 induced or general, where all models achieve F1-score above $0.97$. We can conclude that voice and speech features are strong indicators of RI presence (such as the patient's pause distribution~\citep{fernandessvartman22_speechprosody}), despite not containing much information about the SpO2. As a possibility, we hypothesize that the treatment patients received in the hospitals led to temporary improvement in their SpO2 levels, bringing those closer to healthy levels, but their voice and speech still retains traces found in RI patients. As a result, models can pick up those traces to identify RI but they are not very informative when it comes to the actual SpO2 level.

Models perform better for end-to-end classification than for regression. As SpO2 regression is ineffective, any attempts to do RI classification intermediated by regression are also ineffective.  Lastly, some of the pretrained models using supervised learning are only pretrained for classification tasks, which may lead to poorer performance in downstream regression tasks. While some models use generic pretraining tasks, our experience, as well as that of other researchers, shows that current self-supervision techniques do not translate well for regression. As Transformers are known to perform well in time series forecasting tasks~\citep{wu2020deep} which involve regression, it seems that current self-supervision techniques do not capture dynamic information on how the data space changes and how those changes might affect downstream tasks. If it is possible for the models to learn such information, they might become more effective for regression tasks, which could improve our results for that case. 
Keep in mind, though, that the binary classification task used shows that SpO2 estimation from voice is inherently hard. As a consequence of the binary classification task performance, our results show the limits of using voice and speech as biomarkers.

\section*{Acknowledgements}
This work was supported by FAPESP grants 2022/16374-6 (POSTDOC)  and 2023/00488-5 (SPIRA), and by Coordenação de Aperfeiçoamento de Pessoal de Nível Superior - Brasil (CAPES) - Finance Code 001. Carried out at the Center for Artificial Intelligence (C4AI-USP), supported by FAPESP grant 2019/07665-4 and by the IBM Corporation. The sponsor's involvement in the work was limited to the provided funding. Marcelo Finger was partly supported by CNPq grant PQ 302963/2022-7.





\bibliographystyle{elsarticle-num} 
\bibliography{references}

\end{document}